\documentclass[draft]{agujournal2018}
\usepackage{apacite}
\usepackage{url} 
\usepackage{lineno}
\usepackage{amsmath}

\draftfalse

\journalname{arXiv}

\begin{document}

\title{A novel approach to exploit elastic deformation to constrain regional ice mass change in Antarctica}

%
%




\authors{W.J. Durkin$^{1,2}$, T. Wilson$^{3}$, M. Bevis$^{3}$}

\affiliation{1}{Byrd Polar and Climate Research Center, Ohio State University, Columbus, Ohio, U.S.A}
\affiliation{2}{The MITRE Corporation, Bedford, Massachusetts, U.S.A}
\affiliation{3}{School of Earth Sciences, Ohio State University, Columbus, Ohio, U.S.A}





\correspondingauthor{W.J. Durkin}{wjdurkin.6371@gmail.com}



\begin{keypoints}
	\item Far-field ice loss is pervasive in the elastic deformation at the locations of Antarctic GNSS sites
	
	\item A novel technique to separate regional and far-field deformation is presented
	
	\item Existing distribution of Antarctic GNSS receivers are capable of isolating the elastic deformation due to ice loss of each of Pine Island Glacier, Thwaites Glacier, and Pope-Smith-Kohler glaciers demonstrated using our new technique
	
\end{keypoints}

\begin{abstract}
Across the ANET GNSS network, decadal mass loss of the Antarctic ice sheet drives elastic uplift rates of up to 20 mm~yr$^{-1}$. We explore, for the first time, the viability of using elastic deformation observed in the ANET GNSS network to constrain ice mass changes in Antarctica. We begin by estimating which regions of ice mass loss contribute to the elastic deformation observed at each ANET GNSS site. This is done using an observer-centric model of elastic deformation, which we call an Load Sensitivity Kernel (LSK). Centering the LSK at a GNSS site and scaling by the decadal rate of ice mass loss expresses how much each unit area of mass loss contributes to the elastic deformation observed at the GNSS site. The elastic deformation at the majority of ANET sites is dominated by far-field mass change: 70\% of sites require ice mass change within a radius of 200km or greater to be included before 90\% of the site’s elastic deformation is recovered. We show that the difference between LSKs of a pair of neighboring GNSS receivers can be used to localize the near-field mass change responsible for the differential displacement. Finally, we generalize this in a novel approach that uses an arbitrary number of GNSS sites. In this approach, LSKs of each site are weighted and added in a linear series. These weights are optimized such that the elastic deformation from mass change outside a Region of Interest is minimized. We demonstrate that 7 ANET GNSS sites in the Amundsen Sea region can isolate the elastic deformation, in both the vertical and horizontal components, that results exclusively from PIG's mass loss and/or the losses of the PSK glaciers.  With the current GNSS receiver distribution, the east component of elastic deformation due to mass change at Thwaites can be expressed relative to mass changes at PSK.  Our results demonstrate that the current ANET configuration can use elastic deformation to constrain ice mass changes in Antarctica, especially in regions with relatively dense clusters of GNSS.   
\end{abstract}

\section{Introduction}

Changes in loading on the Earth's surface induce a nearly instantaneous elastic deformational response \citep{farrell1972deformation}. As the timescales of changes in the surface load increase to decadal or longer, the Earth begins to respond increasingly with delayed viscoelastic deformation due to flow of material in the mantle \citep[e.g., glacial isostatic adjustement;][]{nield2014rapid, ivins2020linear}. At shorter timescales, for instance multiannual or less, it is often sufficient to approximate the Earth's deformation as purely elastic \citep[e.g.,][]{bevis2012bedrock,liu2017annual}. Annual rates of elastic deformation in response to climate induced deglaciation range from the scales of a few millimeters to a few centimeters per year in the vicinity of glaciers and ice sheets. This elastic deformation in response to changes in the cryosphere can be measured precisely with GNSS receivers mounted onto rock outcrops and used to provide valuable insights into ice mass change and dynamic glacier processes \citep[e.g.,][]{hansen2021estimating,compton2017short}. Satellite altimetry, imagery, and gravimetry have repeat coverage on timescales of several days to weeks, and the the daily solutions of ground deformation measured by GNSS receivers can provide a highly valuable, dense timeseries of mass change to complement satellite observations. GNET, a network of over 50 GNSS receivers in Greenland, has been used in several studies to relate elastic deformations to changes of the Greenland ice sheet as well as specific Greenland outlet glaciers \citep[e.g.,][]{adhikari2017mass,hansen2021estimating,khan2007elastic,nielsen2013vertical}. 

Between 2003-2018, the Antarctic ice sheet lost an average of $\sim$118~Gt~yr$^{-1}$, with mass gains in East Antarctica partially compensating losses of $\sim$-39~Gt~yr$^{-1}$ in the Antarctic Peninsula and $\sim$169~Gt~yr$^{-1}$ in West Antarctica \citep{smith2020pervasive}. ANET, an Antarctic network of GNSS receivers, records the elastic deformation due to the ongoing loss of the Antarctic ice sheet with amplitudes of up to 2~cm~yr$^{-1}$ in the vertical and $\sim$0.5~cm~yr$^{-1}$ in the horizontal \citep{durkin2020reevaluating} However, there has not yet been a study that has used this as a source of signal to investigate ongoing changes of the Antarctic ice sheet similar to what has been done in Greenland. Although many studies have modeled the elastic deformation of the Antarctic continent due to ongoing ice loss, these are primarily focused on removing this signal from the GNSS receiver observations to leave the residual, inferred as viscoelastic deformation of glacial isostatic adjustment, for the purpose of placing constraints on mantle rheology \citep[e.g.,][]{nield2014rapid,barletta2018observed,martin2016assessment}. Between 2010-2014 and 2015-2018, concentrated regions of rapid thinning ($>$1.5~m~yr$^{-1}$) near the northern flanks of Pine Island Glacier underwent a shift their positions by 10's of kilometers and a $\sim$50\% change in amplitudes \citep{bamber2020complex}. \citet{bamber2020complex} suggested that the proximity of these previously unnoticed rapidly thinning regions to INMN, a GNSS site located $~$50~km away, could bias estimates of the site's elastic uplift rates if not accounted for, leading to similar biases in inferences of the site's viscoelastic deformation and ultimately increased uncertainty in the region's mantle rheology \citep[e.g.,][]{barletta2018observed}. If the elastic deformations at the locations of INMN or other ANET GNSS sites are indeed sensitive to the types of changes in ice thinning noted by \citet{bamber2020complex}, they could represent an untapped dataset in monitoring the changes in Antarctic glaciers similar to what has been done in Greenland \citep[e.g.,][]{hansen2021estimating}. 

In this study, we investigate if elastic deformation at the current distribution of ANET GNSS receivers in Antarctica are sufficient for constraining mass changes of specific Antarctic glaciers and other regions of interest on the continent. Because the elastic deformation fields of many sources of mass change superimpose on one another, it is often difficult to isolate the deformation due to the mass change that is of interest \citep{wahr2013use}. Load induced elastic deformation fields are also capable of covering great distances across the Earth \citep{ludwigsen2020vertical,coulson2021global}. The deformation due to far-field mass change can be a pernicious source of bias if not accounted for. One method to mitigate this is to model and remove these deformation fields \citep[e.g.,][]{compton2017short}. Another method of isolating the deformation of near-field sources of interest from that of far-field mass change is to consider deformation at GNSS sites located very close (within a few 10's of kilometers) from the near-field source of ice loss \citep[e.g.,][]{adhikari2017mass,bevan2015seasonal}. A third approach that has been used is to consider differences in deformation between proximal pair of GNSS sites so that common sources of field deformations largely negate \citep[e.g.,][]{wahr2013use,nielsen2013vertical,liu2017annual}. To understand how the ANET GNSS receiver network responds to the elastic deformation of different near- and far-field ice loss sources, we identify the sources of elastic deformation for individual GNSS, pairs of GNSS, and finally with a novel approach that makes use of the relative deformation among collections of many GNSS.

We find that dynamic ice losses in Antarctica are capable inducing elastic deformation fields with spatial scales of over several hundreds of kilometers and that far-field ice loss constitutes a significant portion of the elastic deformation of over 70\% of ANET GNSS sites. Attempts to constrain the mass change of specific Antarctic glaciers using individual GNSS sites are likely to be poorly constrained in many regions of the continent. While considering the difference between pairs of proximal GNSS sites does narrow the region of contributing ice loss, we find that differential elastic deformation can still exhibit significant sensitivity to dynamic ice loss occurring several 100's of kilometers away in many settings. Using our novel approach that compares the relative deformation among collections of GNSS, we demonstrate that the current ANET GNSS network is capable of isolating the elastic deformation resulting from losses of Pine Island Glacier, the Pope Smith Kohler glaciers, and capable of improving constraints on the losses near the grounding line of Thwaites Glacier. While our focus is primarily on three major glaciers in the Amundsen Sea region, the results and methods presented here are relevant to other places in Antarctica any any other region exhibiting load-induced elastic deformation where there are clusters of GNSS stations.

\begin{figure}
	\centering
	\includegraphics[width=1\textwidth]{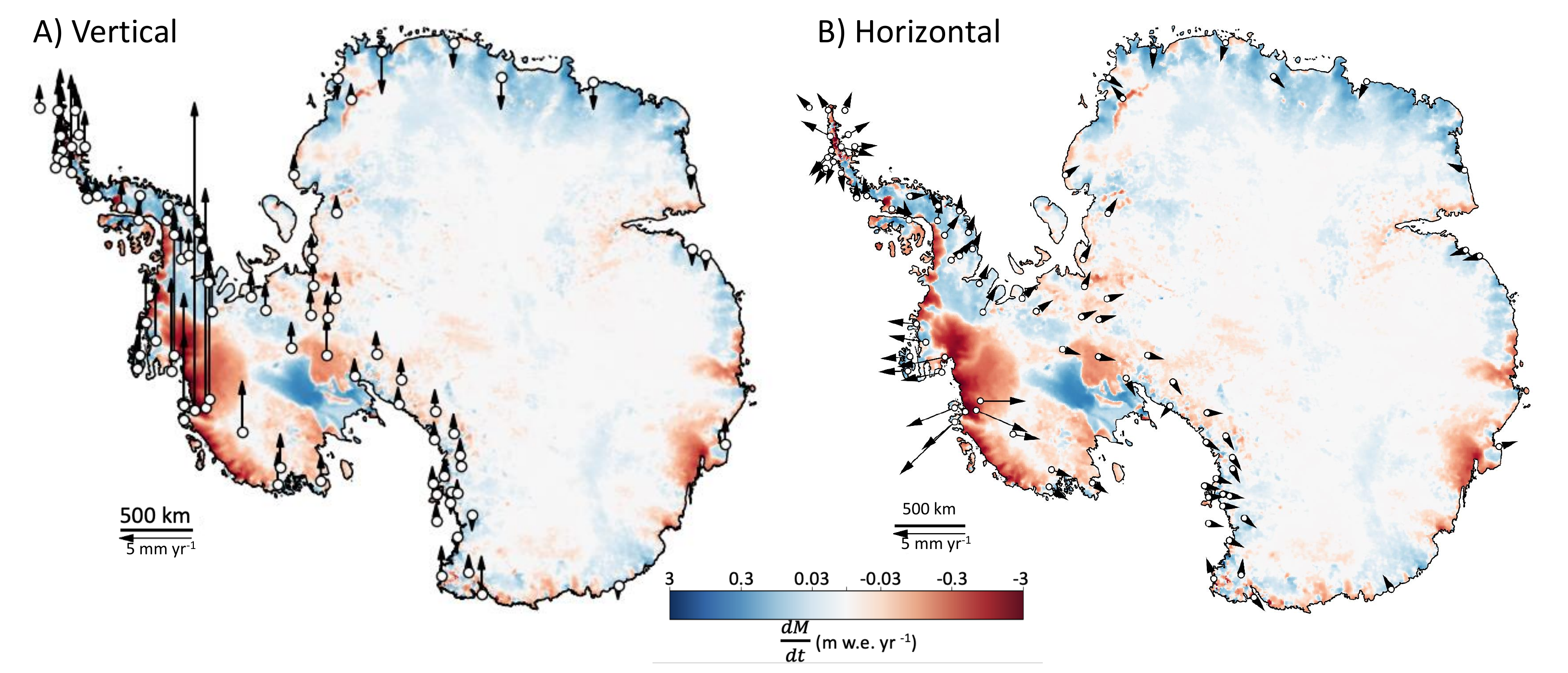}
	\caption{Rate of ice mass change ($\frac{dM}{dt}$) of the Antarctic ice sheet between 2008-2018. $\frac{dM}{dt}$ is based on the altimetry timeseries of Schroeder et al. (2018) and the firn densification model of Medley et al. (2020). Black arrows show the vertical (A) and horizontal (B) components of ice-induced elastic deformation. In region's such as the Weddell Sea, the direction of horizontal elastic deformation is the result of mass change several hundreds of kilometers away located in separate ice drainage basins.}
	\label{fig:dmdt_vector}
\end{figure}

\section{Data and Methods}
\subsection{Decadal Rate of Ice Mass Loss}
To model elastic deformation rates in Antarctica, we first estimate the rate of the ice sheet's change in elevation ($\frac{dh}{dt}$) between 2008-2017.9 using timeseries of satellite altimetry \citep{schroder2019four} augmented with digital elevation map (DEM) based rates of elevation change in select regions \citep{rott2018changing}. Rates of mass change ($\frac{dM}{dt}$) are estimated by using a density of 917~kg~m$^{-3}$ following the removal of contributions to $\frac{dh}{dt}$ from solid Earth deformation and rate of change in firn air content \citep{medley2020forty}. 

Volume change rates are derived from the gridded satellite-altimetry timeseries of \citet{schroder2019four}. In total, \citet{schroder2019four}'s gridded timeseries is constructed from the datasets of 7 satellite altimetry missions corrected for inter-mission offsets and spans the years 1978-2017.9. To estimate the current average decadal rate of Antarctic ice loss, we use elevation measurements between the years 2008 and 2017.9. Ice elevations in this portion of the timeseries were estimated based on Envisat, ICESat, and CryoSat-2 altimetry and expressed on a 10~km resolution grid with monthly timesteps \citep{schroder2019four}. We estimate the $\frac{dh}{dt}$ by fitting a linear trend to the elevation timeseries on a pixel-by-pixel basis \citep[e.g.,][]{willis2012ice}. A gap in $\frac{dh}{dt}$ surrounding the south pole, sometimes referred to as the ''pole hole", exists due to the limitations in coverage imposed by the inclination of the polar orbits of the satellites used in constructing the elevation timeseries. We use a nearest neighbor approach to interpolate this gap in $\frac{dh}{dt}$. 

The altimetry-based $\frac{dh}{dt}$ poorly samples the rapidly thinning outlet glaciers in the northern Antarctic Peninsula due to the glaciers' steep topography and narrow geometries \citep{schroder2019four}. We express the $\frac{dh}{dt}$ of the outlet glaciers that drain into the Larsen A and B embayments of the northern Antarctic Peninsula using $\frac{dh}{dt}$ derived from Digital Elevation Models (DEMs) that span the years 2011-2013 and 2013-2016 \citep{rott2018changing}. Thinning rates during 2011-2016 are estimated by taking the average of DEM-$\frac{dh}{dt}$ grids weighted by the fraction of the 5 year period that they cover. The 2011-2016 thinning rates of glaciers flowing into the Larsen A and B embayments are assumed to be equal to their decadally averaged thinning rates, and the DEM-based $\frac{dh}{dt}$ is incorporated into the Antarctic-wide $\frac{dh}{dt}$ grid. 

This $\frac{dh}{dt}$ grid expresses the sum of elevation changes due to ice change (from both surface mass balance and dynamic thickening/thickening of the ice column), changes in the firn air content, and viscoelastic solid Earth deformation \citep{medley2020forty}. In order to estimate the rate of mass change of the Antarctic ice sheet, we must first remove the contributions of solid Earth deformation and firn air compaction from the $\frac{dh}{dt}$. Solid Earth uplift rates are the combination of the delayed viscous response to past ice changes (i.e., glacial isostatic adjustment) and the immediate elastic response to present-day ice changes \citep[e.g.,][]{martin2016assessment}. The contributions of glacial isostatic adjustment are removed from the $\frac{dh}{dt}$ by subtracting uplift predicted by the W12 GIA model \citep{whitehouse2012new}. Following this, we scaled the residual $\frac{dh}{dt}$ by a factor of 1.0205 as a simple approximation to account for ice-induced elastic deformation rates \citep[e.g.,][]{schroder2019four}. The elevation changes due to the firn air content's rate of change ($\frac{dFAC}{dt}$) is estimated for the period 2008-2017.9 using the Community Firn Model \citep{medley2020forty}. Once the $\frac{dFAC}{dt}$ has been removed from the $\frac{dh}{dt}$, $\frac{dM}{dt}$ is estimated by using a density of 917~kg~m$^{-3}$. Uncertainties in the density model and volume change estimates are not considered in this study, as we are primarily interested in characterizing the geographic sources of elastic deformation of each GNSS receiver, and this will not be majorly affected by density uncertainties. 

\subsection{Elastic Deformation Modeling}
Elastic deformation modeling is performed using the Regional ElAstic Rebound calculator  \citep[REAR;][]{melini2015rebound}. REAR models the elastic deformation using a 1D, spherical, layered Earth. We are using the Preliminary Reference Earth Model (PREM), with a crustal layer modified to have rigidity properties of continental crust \citep{dziewonski1981preliminary}. The load Love numbers are computed to a harmonic degree of 40,000 using \texttt{giapy} \citep{kachuck2019benchmarked}. This harmonic degree is sufficient for modeling elastic deformation with spatial scales of 1~km or greater \citep{bevis2016computing}. The annual rates of ice-induced elastic deformation for each of the 89 ANET sites are summarized in Table S1.  

\subsubsection{Discretization of $\frac{dM}{dt}$}
We compute elastic deformation by convolving a Green's Function with a disc load \citep{melini2015rebound}. When this is performed over a large region, such as the Antarctic Ice Sheet, it is greatly simplified by expressing the mass change as a collection discs of equal radii.  To facilitate this, we first sample the $\frac{dM}{dt}$ onto a hexagonal grid using the icosahedron-based H3 discrete global gridding system \citep[DGGS;][]{brodsky2018h3}. The hexagonal grid of H3 allows for closer packing of the surface load discs than does a square grid. The fraction of the hexagonal cell uncovered by the disc inscribed in it is $\sim$10\%, whereas the fraction uncovered by a disc inscribed in a square cell is $\sim$27\%. As a result of the distortion from H3's map projection, areas of the hexagonal cells in H3 vary by up to $\pm$24\% from their reported average area. We sample the $\frac{dM}{dt}$ onto a hexagonal grid in which each cell has roughly the same area as a 6~km square (H3 resolution R6). To account for distortion of the H3 gird, the $\frac{dM}{dt}$ is scaled by the ratio of the hexagonal cell's area over the mean hexagonal cell area of that resolution. We account for the missing space between discs by scaling each disc's $\frac{dM}{dt}$ by the ratio of the hexagonal cell's area to the area of the inscribed disc (a factor of $\frac{2\sqrt{3}}{\pi}$, or roughly 1.10). 

\begin{figure}
	\centering
	\includegraphics[width=0.9\textwidth]{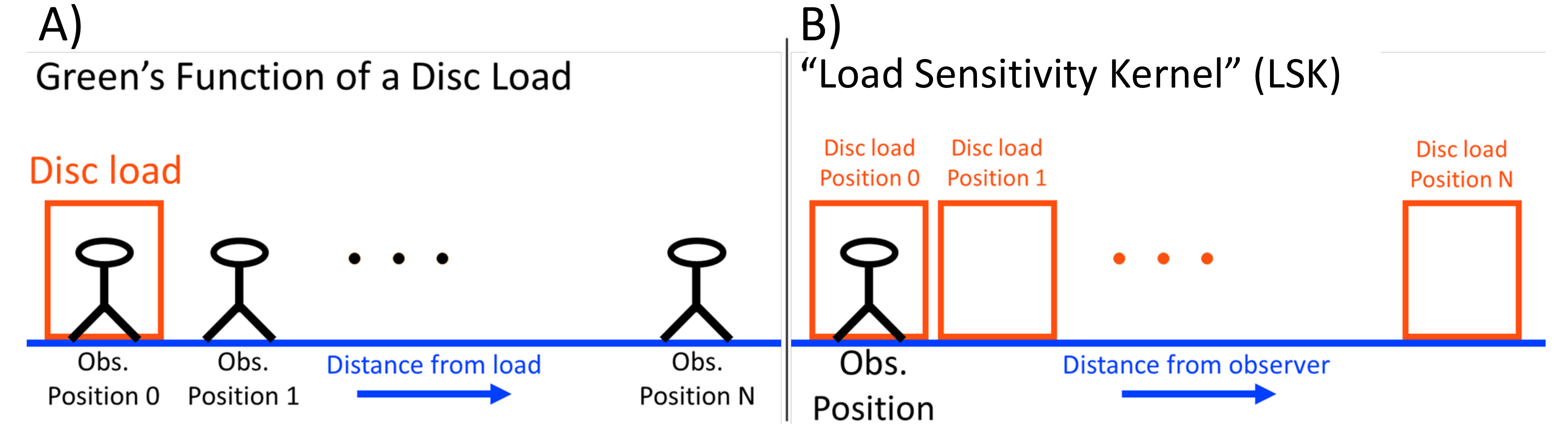}
	\caption{A) An illustration of the disc-centric Green's function used in elastic deformation modeling. A unit disc load is placed at the origin, and elastic displacement is modeled for all distances from the disc's center. B) A Load Sensitivity Kernel (LSK) depicting an observer-centric description of elastic deformation. An observer is placed at the origin, and elastic deformation at the origin is computed for a disc located an arbitrary distance away.}
	\label{fig:igf_cartoon}
\end{figure}

\subsubsection{Load Sensitivity Kernels}
In order to map the geographic origins of ice loss contributing to the elastic deformation observed at a GNSS site \citep[e.g.,][]{agnew2001map}, we expand on a technique used by \citet{adhikari2017mass} to compute the sensitivity gradient of elastic deformation. We refer to this sensitivity gradient as a Load Sensitivity Kernel (LSK). Elastic deformation is often modeled using a disc-centric approach in which an elastic Green's Function is placed at the center of a disc and deformation is computed for any arbitrary distance and angle from the disc's center. This is then performed for all discs, and the total elastic deformation is computed by superimposing the deformation of all disc loads. Instead, we take an observer-centric approach that expresses the deformation observed at a fixed observation site that results from a disc located at an arbitrary distance from the observer (Figure~\ref{fig:igf_cartoon}). When a LSK is centered at a GNSS site, it expresses the contribution of a unit disc load at an arbitrary distance and angle away to the elastic deformation observed at the GNSS site. Multiplying the GNSS-centered LSK by the $\frac{dM}{dt}$ maps the contribution of each area of ice unloading to the elastic deformation at the GNSS site.  

The LSK is computed numerically by placing a 1 meter water equivalent (m~w.e.) disc with a radius of 3.23~km (i.e., a disc inscribed in a R6 hexagon) at 0$^{o}$~N, 0$^o$~E and computing the elastic deformation at the origin. The elastic deformation at  0$^{o}$~N, 0$^o$~E is computed again after translating the disc east a distance of $\frac{1}{3}$ of its radius. This is repeated until the disc is located 5000~km away from the origin, and the elastic displacement is interpolated al ong this line using a cubic spline. We then center the resulting LSK on one of the ANET GNSS sites (e.g., INMN shown in Figure~\ref{fig:igf_inmn}~A-C) and multiply by the $\frac{dM}{dt}$ (Figure~\ref{fig:igf_inmn}~D-F). This expresses the contribution of each $\frac{dM}{dt}$ disc to the elastic deformation signal observed at the GNSS site. This is repeated for all 89 of the ANET sites.

\section{Results and Discussion}
\subsection{Pervasive Influence of Far-Field Ice Loss} 
\begin{figure}[h]
	\centering
	\includegraphics[width=1\textwidth]{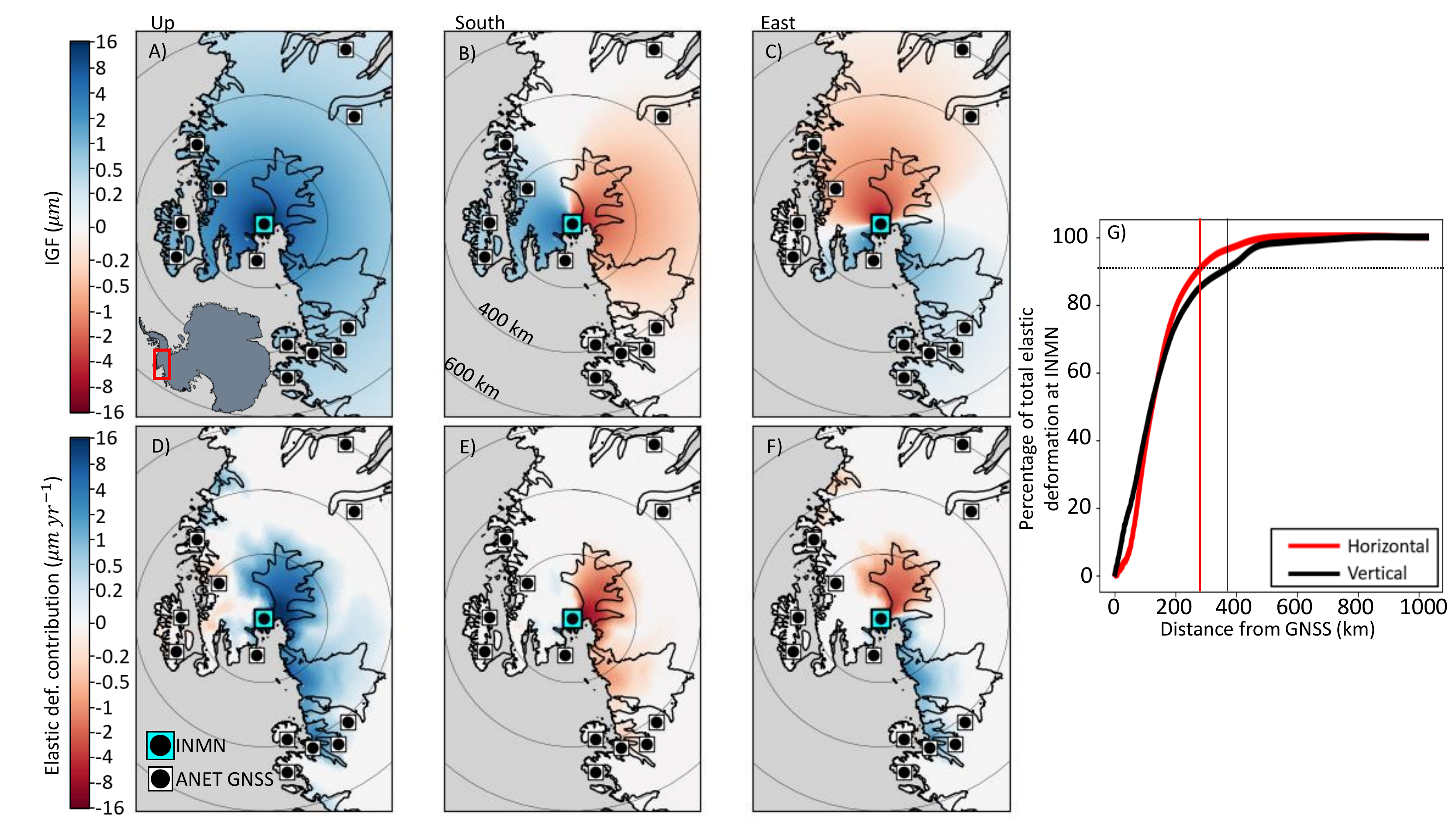}
	\caption{A-C) A Load Sensitivity Kernel (LSK) centered at the INMN GNSS site in the Amundsen Sea region. The LSK depicts the displacement at INMN resulting from a unit load at an arbitrary location. D-F) The decadal rate of elastic displacement at INMN due to each unit of observed $\frac{dM}{dt}$. G) The deformation contribution of each $\frac{dM}{dt}$ disc (D-F) integrated over distance as a percentage of the total elastic deformation rate occurring at INMN}
	\label{fig:igf_inmn}
\end{figure}

We begin by quantifying the origins of ice-induced elastic deformation signals observed at each of the ANET GNSS sites.  To demonstrate this, we first consider the Amundsen Sea region in West Antarctica which is host to prodigious annual ice loss carried out primarily by thinning at the Pine Island, Thwaites, and Pope-Smith-Kohler (PSK) glaciers \citep[e.g.,][]{schroder2019four}. Figure~\ref{fig:igf_inmn}A-D shows an LSK centered at INMN, a GNSS site located $\sim$40~km from the terminus of Pine Island Glacier. Regions with annual horizontal ice flow $>$100~m~yr$^{-1}$ \citep{mouginot2019continent}, a proxy for thinning due to stretching, or dynamic thinning, are outlined in black (Figure~\ref{fig:igf_inmn}). Multiplying $LSK_{INMN}$ by the annual rate of ice mass loss ($\frac{dM}{dt}$, Figure~\ref{fig:dmdt_vector}) yields a map showing the contribution each disc of ice loss contributes to the elastic deformation signal observed at INMN (Figure~\ref{fig:igf_inmn}D-F). 

The INMN site has an elastic deformation rate of 9.4~mm~yr$^{-1}$ in the vertical and 2.2~mm~yr$^{-1}$ in the horizontal (Table S1).This is the result of the sum of many discs, each of which contribute deformation on the order of 0.1 - 10 $\mu$m~yr$^{-1}$

\begin{equation}
\dot\epsilon_{INMN} = \sum_{i=1}^{N} \dot{M}_i(\text{\textbf{LSK}}_{\text{\textbf{INMN}}_i})
\label{eq:lsk_inmn}
\end{equation}

Discs that are closest to INMN with fast thinning rates contribute the most to the site's elastic deformation rate, for example regions near the grounding line of Pine Island Glacier. As the distance from INMN increases, although the contributions of each disc diminishes, the number of discs and increases quite rapidly. 

\begin{figure}[h]
	\centering
	\includegraphics[width=1\textwidth]{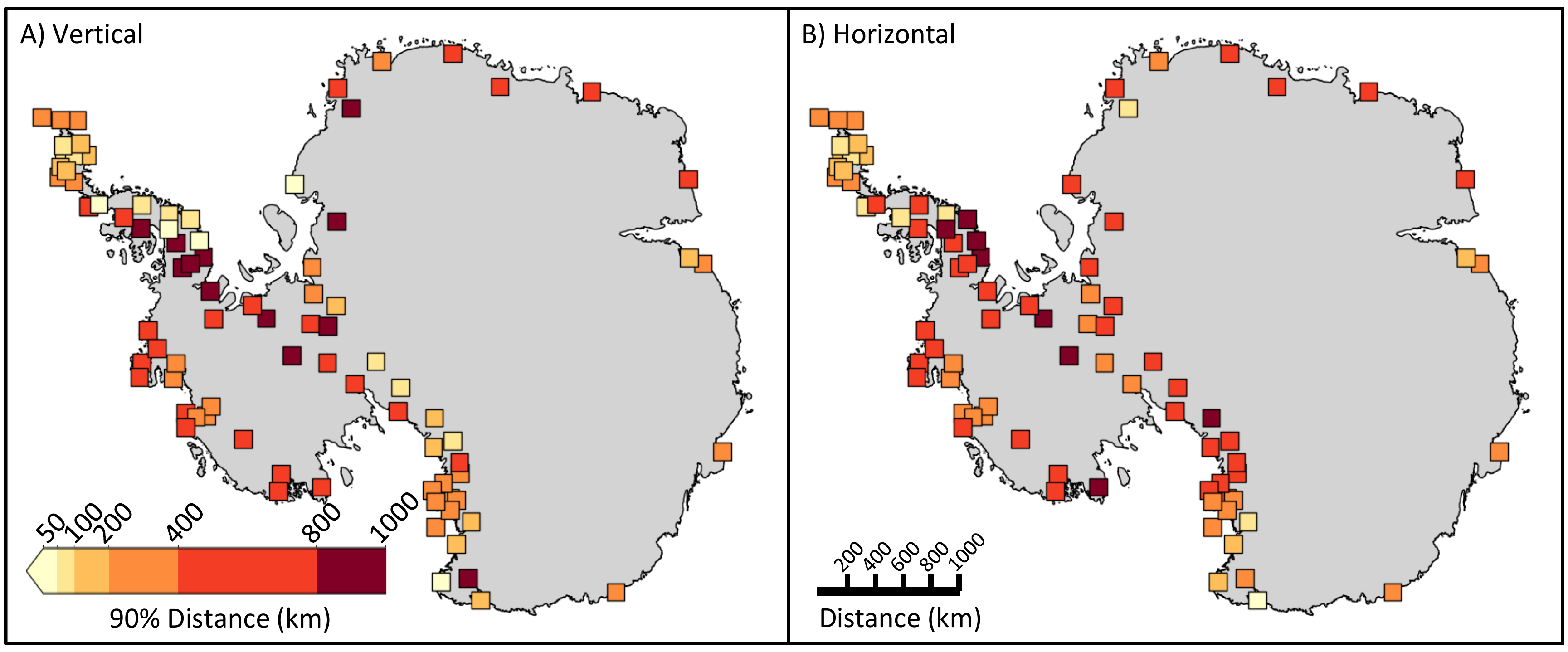}
	\caption{The D$_{90}$, or distance from a GNSS site in which all ice mass change is needed to account for 90\% of the GNSS site's elastic deformation}
	\label{fig:D90}
\end{figure}

By integrating the contributions of each disc in Figure (3D-F) over their distance from INMN (Figure 3G), we find that the deformation due to ice loss in the far-field, here defined as distances $>$200~km, account for roughly half of INMN's elastic deformation rate in both the vertical and horizonal components. To account for 90\% of INMN's elastic deformation, ice loss within a radius of 274~km(horizontal) and 357~km(vertical) must be considered, a distance we refer to as $D_{90}$. When this is repeated at other GNSS sites in the ANET network across the continent, we find that far-field ice loss is a pervasive component in elastic deformation across the continent. Over 70\% of GNSS sites featuring a $D_{90}$ greater than 200~km in both the vertical and horizontal directions (Fig.~\ref{fig:D90}).

In the case of INMN, dynamic losses near the grounding lines of Thwaites and the PSK glaciers contribute as much to INMN's elastic deformation as the upper branches of PIG's dynamically thinning region (Fig.~\ref{fig:igf_inmn}D-F). Indeed, the annual losses of Thwaites and the PSK glaciers constitute 20\% of INMN's elastic uplift rate even though they are located several hundred kilometers away. The sensitivity of INMN to such a large area of ice loss will make it difficult to place constraints on the mass changes individual glaciers of glaciers in the Amundsen Sea region, including the neighboring Pine Island glacier, using the approach \citet{hansen2021estimating} used in Greenland. The strong prevalance of far-field ice loss in the elastic deformation signal of 70\% of ANET GNSS sites indicates that attempts to resolve ice mass changes of specific glaciers or zones of interest using individual GNSS receivers will likely be poorly constrained for most regions in Antarctica. 

Between 2015-2018, concentrated regions of rapid thinning($>$3~m~yr~$^{-1}$) within a 50~km radius of INMN changed by $\sim$50\% with respect to their 2010-2014 thinning rates \citet{bamber2020complex}. \citet{bamber2020complex} suggested that this change in thinning rates could induce a comparable change in INMN's elastic uplift rates, and by extension alter the site's inferred viscous deformation rates and degrade estimates of the region's mantle viscosity. Testing this using out LSK approach shows that the changes in thinning rates of PIG noted by \citet{bamber2020complex} elicit a modest change in INMN's inferred viscous uplift rate that is unlikely to alter interpretations about the region's mantle viscosity. While the mass losses within the 50~km radius from INMN are among the largest contributors per disc load (up to several 10's of microns each), they are relatively few in number. Ice loss within a 50~km radius constitutes $\sim$17\% of INMN's total elastic uplift rate (Fig.~\ref{fig:igf_inmn}). Elastic deformation accounts for roughly 25\% of the site's total uplift rate, with the remainder inferred to be due to viscous deformation associated with glacial isostatic adjustment \citep{barletta2018observed}. A 50\% change in the thinning rates of ice loss in this 50~km radius would alter the inferred viscous deformation by less than 5\%, a change too small to significantly affect interpretations on mantle viscosity. Our results indicate that proximal, concentrated, high-amplitude mass loss expressions of the kind noted by \citet{bamber2020complex} tend not to be dominant features the elastic deformation of ANET GNSS sites, as their high $D_{90}$ values (Fig.~\ref{fig:D90}). It is certainly possible that epochal changes in ice thinning rates are large enough to affect the viscous deformation rates inferred at GNSS sites in the Amundsen Sea region. However, changes over distances of several hundreds of kilometers from the GNSS sites can form a very large percentage of a site's deformation and must be considered to correctly express this (Fig.~\ref{fig:D90}).  

\subsection{Isolating the Deformation due to Mass Change in a Region of Interest}

In the previous section, we demonstrated that far-field ice loss is pervasive in the elastic deformation of Antarctic GNSS sites. In order to use elastic deformation of these GNSS sites to constrain the mass change of individual Antarctic glaciers, it is first necessary to remove these far-field contributions. One approach that has been used to isolate the deformation field of near-field mass change is to consider the difference in elastic deformation between a pair of GNSS sites \citep[e.g.,][]{wahr2013use,nielsen2013vertical,liu2017annual}. 

\begin{figure}
	\centering
	\includegraphics[width=1\textwidth]{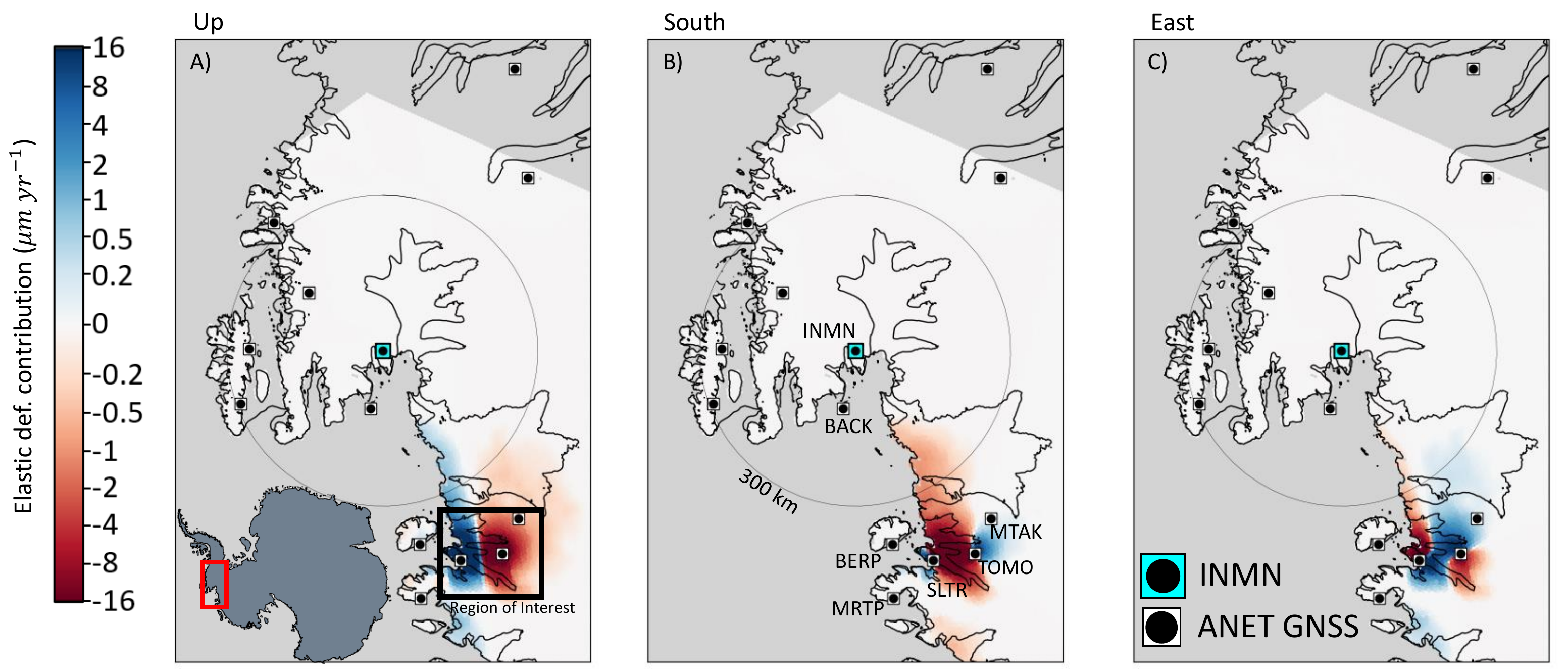}
	\caption{A-C) Differential Load Sensitivity Kernels (dLSKs) formed from the difference between LSKs centered at TOMO and SLTR GNSS sites scaled by $\frac{dM}{dt}$. Each pixel represents the contribution of a disc of ice loss to the differential elastic deformation between TOMO and SLTR. The difference in elastic deformation between this neighboring pair of GNSS sites remains sensitive to ice loss over 300~km away.}
	\label{fig:dLSK_PSK}
\end{figure}

We demonstrate the effectiveness of this approach in isolating the elastic deformation of the PSK glaciers by considering the difference in LSKs (dLSK) of the nearby SLTR and TOMO sites, expressed as

\begin{equation}
\Delta\dot\epsilon = \sum_{i=1}^{N} \dot{M}_i(\text{\textbf{LSK}}_{\text{\textbf{TOMO}}_i} - \text{\textbf{LSK}}_{\text{\textbf{SLTR}}_i})
\label{eq:lsk_diff}
\end{equation}

where $i$ is $i^{th}$ disk of ice mass change. This dLSK approach is shown in Figure ~\ref{fig:dLSK_PSK}. In the vertical and east components, the differential pair is sensitive almost exclusively to mass losses of the PSK glaciers, while the south component also experiences some modest contributions from losses near the grounding line of Thwaites. The PSK glaciers are estimated to contain an ice volume capable of $\sim$1.2~m of global sea level rise \cite{milillo2022rapid}. Although this is modest in comparison to the capacity for future sea level rise of the neighboring Pine Island and Thwaites glaciers, the PSK glaciers are undergoing grounding line retreat and mass loss (>5~m~w.e./yr) at prodigious rates that are expected to be diagnostic of marine ice sheet instability \citep[MISI;][]{schoof2012marine, milillo2022rapid}. A better understanding of MISI at the PSK glaciers would be valuable in predicting how this process may affect the much larger Pine Island and Thwaites glaciers in the future. An additional concern is that the Smith branch of the PSK glaciers may propagate higher thinning rates inland towards the ice divide of Thwaites Glacier and hasten its demise \citep{lilien2019melt}.

The dLSK (Fig.~\ref{fig:dLSK_PSK}) shows that the difference in elastic deformation between the SLTR-TOMO pair results almost exclusively from PSK mass loss in the vertical and east components with modest contributions from mass loss near the grounding line of Thwaites in the south component (Fig. XB). In addition to providing constraints on the magnitude of ice mass loss at the PSK glaciers, the dipole pattern of the dLSK shows that differences between SLTR and TOMO could be used to estimate the location of where ice loss of PSK is occurring. In the case of grounding line retreat or increased  thinning near the grounding line, the difference in elastic deformation of the SLTR TOMO pair would increase in the vertical component and decrease in the south and east components. In the case of thinning that extends inland towards the margin of Thwaites Glacier, this difference would decrease in the vertical, increase in the east component, and have a modest decrease in the south component. 

\begin{figure}[h]
	\centering
	\includegraphics[width=1\textwidth]{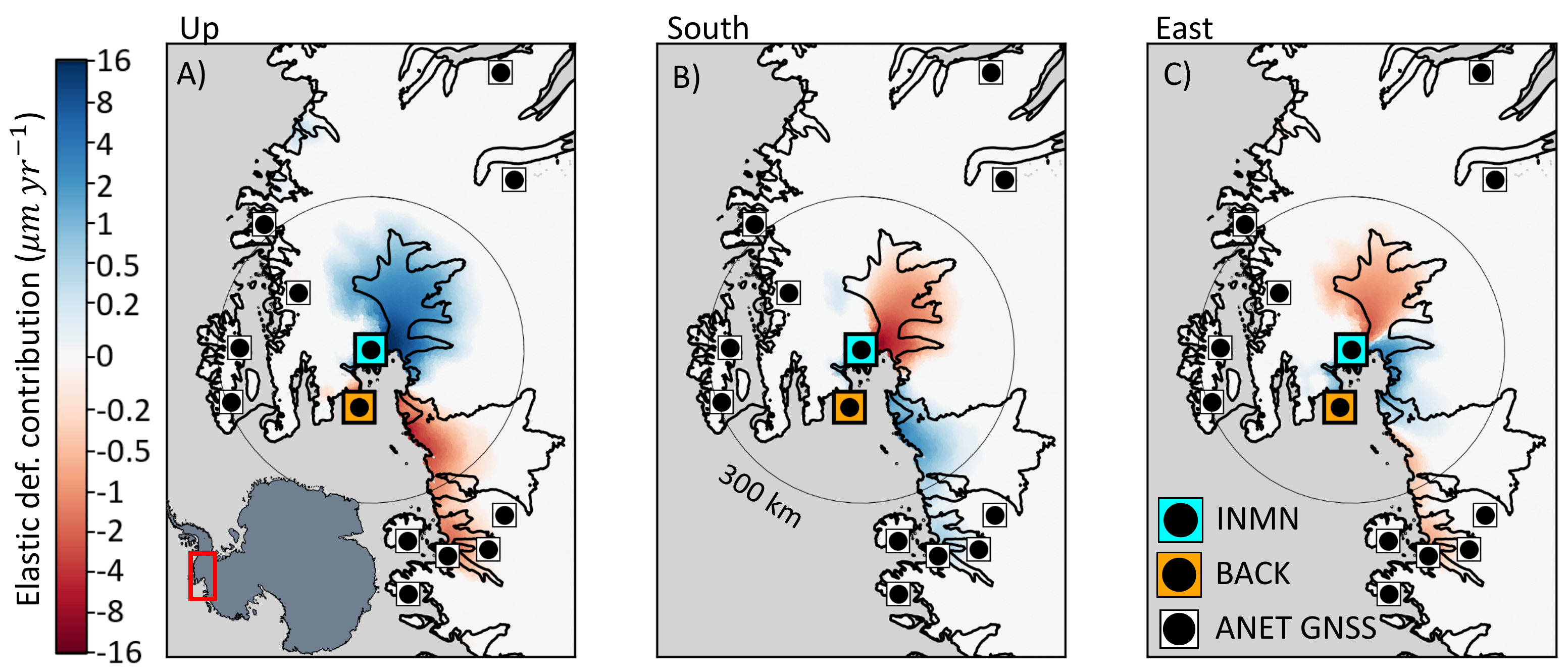}
	\caption{A-C) Differential Load Sensitivity Kernels (dLSKs) formed from the difference between LSKs centered at INMN (cyan) and BACK (orange) GNSS sites scaled by $\frac{dM}{dt}$. Each pixel represents the contribution of a disc of ice loss to the differential elastic deformation between INMN and BACK. The difference in elastic deformation between this neighboring pair of GNSS sites remains sensitive to ice loss over 300~km away. }
	\label{fig:dIGF_inmn_back}
\end{figure}

To test the effectiveness of a differential pair of GNSS sites in isolating the deformation due to mass change of PIG, we consider the difference in elastic deformation of INMN and the nearest GNSS site, BACK, located 115~km away. Contributions from mass loss outside of PIG are largely removed from the easterly component of the INMN/BACK pair due to the general east-west orientation of PIG, Thwaites, and the PSK glaciers. In the vertical and southerly components, although the regions of ice loss that contribute to elastic deformation to the differential pair are more narrowly focused compared to the single GNSS case (Fig.~\ref{fig:dIGF_inmn_back}), far-field mass loss is remains a significant component of the INMN/BACK pair. 

While the INMN/BACK does not completely isolate the mass change of PIG from other sources of mass change, it can be used to express mass changes of PIG relative to changes of Thwaites and the PSK glaciers. An increase in the vertical component of the INMN/BACK difference would indicate an increase in the thinning rates of PIG and/or a decrease in thinning rates of the PSK glaciers and regions near the grounding line of Thwaites glacier. A similar situation exists for the southern component of deformation but with the signs reversed. In the east component, an increase/decrease in the INMN/BACK pair's difference primarily expresses the difference in the mass loss rates near the grounding line of PIG relative to mass losses in its upper branches (Fig.~\ref{fig:dIGF_inmn_back}C). 

To better isolate the elastic deformation that results solely from PIG, we introduce a novel approach that generalizes Eq.~\ref{eq:lsk_diff} to make use of multiple GNSS sites. This is done by considering LSK's centered at many GNSS receiver locations added in a weighted series, expressed as 

\begin{align}\nonumber
\dot\epsilon_{ROI} = \sum_{i=1}^{N} \dot{M}_i(&a\,\, \text{\textbf{LSK}}_{\text{\textbf{INMN}}_i} + b\,\, \text{\textbf{LSK}}_{\text{\textbf{BACK}}_i} + c\,\, \text{\textbf{LSK}}_{\text{\textbf{BERP}}_i} + d\,\, \text{\textbf{LSK}}_{\text{\textbf{TOMO}}_i} \\  
&+ e\,\, \text{\textbf{LSK}}_{\text{\textbf{SLTR}}_i} + f\,\, \text{\textbf{LSK}}_{\text{\textbf{MTAK}}_i} + g\,\, 
\text{\textbf{LSK}}_{\text{\textbf{MRTP}}_i} + \ldots\,\,)
\label{eq:weighted_lsk}
\end{align}

where $\dot\epsilon_{ROI} $ is the total elastic deformation due to ice mass change within a Region of Interest (ROI). 

\begin{figure}[h]
	\centering
	\includegraphics[width=1\textwidth]{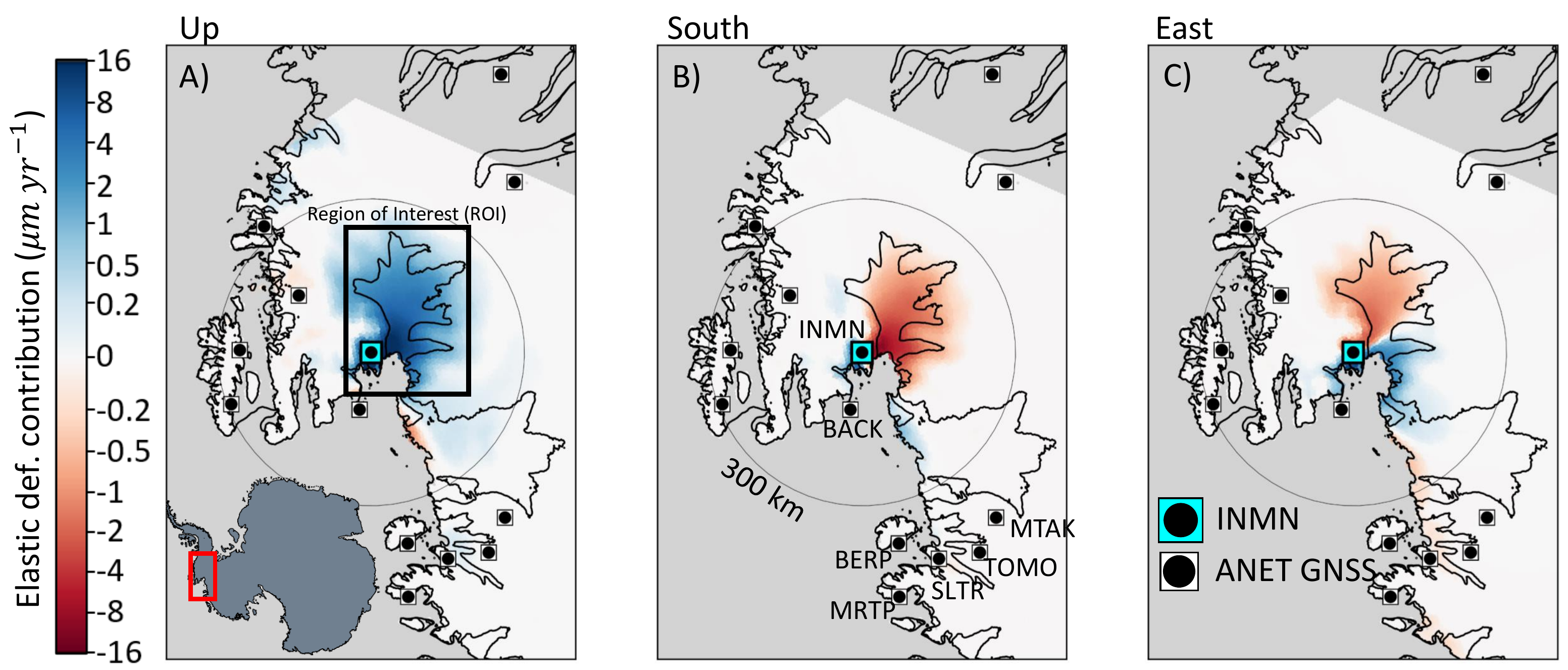}
	\caption{A-C) LSKs centered on the GNSS sites are added in a weighted series (wLSK) and scaled by $\dot{M}$. The weights of each LSK are chosen to minimize the elastic deformation response to mass change outside a Region of Interest (ROI) centered on Pine Island Glacier. The wLSK demonstrates that the elastic deformation signal of Pine Island Glacier can be isolated in the vertical, south, and east components when considering the relative deformation at many ANET GNSS sites in their current distribution.}
	\label{fig:wIGFS_PIG}
\end{figure}

We search for the combination of parameters that minimize the sum of elastic deformation originating from outside of the ROI. To avoid the trivial solution, in which all coefficients are zero, we force the coefficient of the GNSS site nearest to the ROI to be 1. The remaining coefficients are bound between $\pm$1. Our optimization is performed using the iterative, modified global CORS method of \citet{knysh2016blackbox}. For each iteration, the cost function is sampled using a Latin Hyper cube. The cost function surface is then estimated by interpolating these samples using a radial basis function, and the region surrounding the estimated minima are explored in subsequent iterations. This is performed until convergence or the completion of 100,000 trials.

The weighted LSK series (wLSK) focused on PIG is shown in Figure~\ref{fig:wIGFS_PIG}, and the weights are reported in Table~S1 of the supplementary material. This approach demonstrates that the relative deformation among the ANET GNSS receiver locations in their current configuration are capable of isolating elastic deformation caused by the mass loss of PIG. In the vertical and south components, the uniform sign of the wLSK within PIG shows that the glacier system's mass change could be constrained using deformation in these directions, but not the location of the mass change within PIG. In the east component, retreat at the grounding line would result in an increase in the wLSK, while propagation of thinning up stream, into tributary branches, or into slow flowing regions \citep[e.g.,][]{bamber2020complex} would diminish the weighted LSK. This approach is also able to isolate the elastic deformation in the east component caused by thinning near the grounding line of Thwaites Glacier relative to thinning of the PSK glaciers (Supplementary Figure 1)

\citet{ludwigsen2020vertical} and \citet{coulson2021global} suggest removing effects of unwanted far-field ice mass change by modeling their elastic deformation fields and removing from the GNSS signals prior to analysis. We propose that, alternatively, the weighted LSK series approach could be used to remove these global fingerprints of ice-induced elastic deformation. This would allow one to remain agnostic about the ice mass change outside of the study region and avoid potentially time consuming exercise of modeling global elastic deformation.

\section{Conclusions}
We have investigated the ability of elastic deformation at the current distribution of Antarctic GNSS to place constraints on the mass change of individual glaciers in Antarctica, in particular the rapidly diminishing Pine island, Thwaites, and Pope-Smith-Kohler glaciers in the Amundsen Sea region. Far-field ice loss is a pervasive component of the deformation field at Antarctic GNSS sites. Over 70\% of sites require ice loss at distances of $>$200~km to be considered in order to express 90\% of their elastic deformation. Even when a GNSS site is within a few 10's of kilometer of a major center of mass loss (e.g., INMN and Pine Island Glacier, or SLTR and the Pope-Smith-Kohler glaciers), they often still derive a significant portion of their elastic deformation from dynamic ice loss 100's of kilometers away. The strong presence of far-field ice loss signals will likely make attempts to use individual GNSS sites to estimate ice mass change at individual glaciers poorly constrained in most regions of the continent. One exception to this is the North Antarctic Peninsula, where $D_{90}$ distances are consistently less than 200~km in both the horizontal and vertical components of deformation. This is likely due to the region's great distance from the Amundsen Sea region and the presence of narrow outlet glaciers with rapid thinning rates \citep[$>$10~m~yr$^{-1}$;][]{rott2018changing}. 

Using our LSK approach, we have demonstrated that using the difference between a pair of GNSS sites in the Amundsen Sea region capable of isolating the elastic deformation resulting from mass loss of the PSK glaciers and expressing the mass changes of Pine Island glacier relative to the changes of Thwaites and the PSK glaciers. For the PSK glaciers, our differential LSK approach indicates the difference in elastic deformation between TOMO and SLTR is capable of distinguishing between ice loss at the grounding line versus the propagation of ice loss towards the margin of the much larger Thwaites Glacier. For Pine Island Glacier, the east component of differences between INMN and BACK is sensitive almost exclusively to PIG and can resolve changes near the grounding line relative to changes in the upper branches of PIG's dynamically thinning regions.
Using a new approach that makes use of numerous GNSS sites added in a weighted series (wLSK's), we demonstrate that the elastic deformation resulting exclusively to PIG's mass change can be fully isolated in all three deformation components. Using this same approach, mass changes near the grounding line of Thwaites Glacier can be expressed relative to the mass loss of the PSK glaciers.  In our study, we have only tested the simple geometry of a rectangular box to define our region of interest. By modifying the shape of the region of interest, it may be possible to achieve improved isolation of Thwaite's elastic deformation signal with the current GNSS distribution in Antarctica. This approach has applications for isolating ice mass change in the deformation signals of GNSS sites in other regions of Antarctica and is especially applicable where GNSS sites are clustered in the vicinity of changing ice masses.

\acknowledgments
We would like to thank Dr. Ramina Ghods for her patient and constructive comments that have improved this manuscript. This work was funded by the Byrd Polar and Climate Research Center's Postdoctoral Fellowship, the National Science Foundation Grants 0632322, 1249631, and the Ohio State University Grant 1745074. William J. Durkin's affiliation with The MITRE Corporation is provided for identification purposes only, and is not intended to convey or imply MITRE's concurrence with, or support for, the positions, opinions, or viewpoints expressed by the author. Approved for Public Release. Public Release Case Number 22-3449


%

 \bibliography{far-field_sources}
%




\end{document}


%
%

\begin{article}

\title{Supporting Information for "A novel approach to exploit elastic deformation to constrain regional ice mass change in Antarctica"}

\authors{W.J. Durkin$^{1,2}$, T. Wilson$^{3}$, M. Bevis$^{3}$}

\affiliation{1}{Byrd Polar and Climate Research Center, Ohio State University, Columbus, Ohio, U.S.A}
\affiliation{2}{The MITRE Corporation, Bedford, Massachusetts, U.S.A}
\affiliation{3}{School of Earth Sciences, Ohio State University, Columbus, Ohio, U.S.A}


\footnotetext{William J. Durkin's affiliation with The MITRE Corporation is provided for identification purposes only, and is not intended to convey or imply MITRE's concurrence with, or support for, the positions, opinions, or viewpoints expressed by the author. Approved for Public Release. Public Release Case Number 22-3449}


 ------------------------------------------------------------------------ 
\end{article}
\clearpage


%
%

\begin{center}
	\begin{longtable}{|l	|	c	|	c	|	c	|	c	|	c	|	c	|	c	|}
		\caption[The names, locations, and elastic deformation rates in the vertical, south, and east components at the GNSS reciever sites. D$_{90}$ expresses the radius surrounding the GNSS receiver site for which all ice mass must be considered to account for 90\% of the site's elastic deformation.]{The names, locations, and elastic deformation rates in the vertical, south, and east components at the GNSS reciever sites. D$_{90}$ expresses the radius surrounding the GNSS receiver site for which all ice mass must be considered to account for 90\% of the site's elastic deformation.} \label{tab:gnss_coords} \\
		
		\hline \multicolumn{1}{|c|}{\textbf{Site}} & \multicolumn{1}{c|}{\textbf{Lon}} &
	    \multicolumn{1}{c|}{\textbf{Lat}} &
	    \multicolumn{1}{c|}{\textbf{U}} &
	    \multicolumn{1}{c|}{\textbf{S}} &
	    \multicolumn{1}{c|}{\textbf{E}} &
	    \multicolumn{1}{c|}{\textbf{Vert. D$_{90}$}} &
		 \multicolumn{1}{c|}{\textbf{Horiz. D$_{90}$}} \\
		 \multicolumn{1}{|c|}{\textbf{}} &
		 \multicolumn{1}{c|}{\textbf{}} &
		 \multicolumn{1}{c|}{\textbf{}} &
		 \multicolumn{1}{c|}{\textbf{mm~yr$^{-1}$}} &
		 \multicolumn{1}{c|}{\textbf{mm~yr$^{-1}$}} &
		 \multicolumn{1}{c|}{\textbf{mm~yr$^{-1}$}} &
		 \multicolumn{1}{c|}{\textbf{km}} &
		 \multicolumn{1}{c|}{\textbf{km}}  \\ 
		 \hline 
		\endfirsthead
		
		\multicolumn{8}{c}%
		{{\bfseries \tablename\ \thetable{} -- continued from previous page}} \\
		\hline \multicolumn{1}{|c|}{\textbf{Site}} & \multicolumn{1}{c|}{\textbf{Lon}} &
		\multicolumn{1}{c|}{\textbf{Lat}} &
		\multicolumn{1}{c|}{\textbf{U}} &
		\multicolumn{1}{c|}{\textbf{S}} &
		\multicolumn{1}{c|}{\textbf{E}} &
		\multicolumn{1}{c|}{\textbf{Vert. D$_{90}$}} &
		\multicolumn{1}{c|}{\textbf{Horiz. D$_{90}$}} \\
		\multicolumn{1}{|c|}{\textbf{}} &
		\multicolumn{1}{c|}{\textbf{}} &
		\multicolumn{1}{c|}{\textbf{}} &
		\multicolumn{1}{c|}{\textbf{mm~yr$^{-1}$}} &
		\multicolumn{1}{c|}{\textbf{mm~yr$^{-1}$}} &
		\multicolumn{1}{c|}{\textbf{mm~yr$^{-1}$}} &
		\multicolumn{1}{c|}{\textbf{km}} &
		\multicolumn{1}{c|}{\textbf{km}}  \\ 
		\hline 
		\endhead
		
		\hline \multicolumn{8}{|r|}{{Continued on next page}} \\ \hline
		\endfoot
		
		\hline \hline
		\endlastfoot
		
		ABOA & 346.593 & -73.044 & -0.63 & 0.12 & 0.11 & 473 & 708 \\
		BACK & 257.522 & -74.43 & 5.84 & -1.78 & 0.03 & 378 & 343 \\
		BEAN & 290.698 & -75.956 & 0.28 & 0.03 & 0.33 & 971 & 817 \\
		BELG & 325.373 & -77.875 & 0.48 & 0.0 & 0.15 & 11 & 544 \\
		BENN & 243.54 & -84.786 & 1.89 & 0.13 & -0.1 & 713 & 397 \\
		BERP & 248.115 & -74.546 & 7.36 & -2.44 & 0.02 & 421 & 324 \\
		BREN & 296.974 & -72.673 & -0.18 & -0.16 & 0.1 & 73 & 97 \\
		BRIP & 158.469 & -75.796 & -0.19 & -0.03 & -0.02 & 176 & 51 \\
		BUMS & 174.499 & -85.961 & 0.6 & -0.04 & -0.02 & 81 & 757 \\
		BURI & 155.894 & -79.147 & 0.1 & -0.07 & -0.0 & 223 & 482 \\
		CAPF & 299.442 & -66.012 & 1.59 & 0.41 & 0.25 & 156 & 165 \\
		CAS1 & 110.52 & -66.283 & 0.93 & -0.23 & -0.17 & 381 & 336 \\
		CLRK & 218.126 & -77.34 & 1.49 & 0.14 & -0.42 & 783 & 775 \\
		COTE & 161.998 & -77.806 & -0.74 & -0.01 & -0.01 & 111 & 675 \\
		CRDI & 306.801 & -82.862 & 0.83 & -0.03 & 0.2 & 285 & 740 \\
		DAV1 & 77.973 & -68.577 & -0.18 & 0.14 & 0.0 & 213 & 216 \\
		DEVI & 161.977 & -81.477 & 0.33 & -0.05 & -0.05 & 129 & 581 \\
		DUM1 & 140.002 & -66.665 & -0.05 & 0.1 & -0.02 & 244 & 287 \\
		DUPT & 297.183 & -64.805 & 3.67 & -0.32 & -1.19 & 95 & 100 \\
		FALL & 216.368 & -85.306 & 0.31 & -0.15 & -0.32 & 410 & 293 \\
		FIE0 & 168.424 & -76.145 & 0.25 & -0.03 & -0.07 & 306 & 326 \\
		FLM5 & 160.271 & -77.533 & 0.01 & -0.02 & -0.03 & 254 & 372 \\
		FONP & 298.353 & -65.245 & 4.3 & 1.24 & 0.28 & 78 & 74 \\
		FOS1 & 291.679 & -71.313 & 0.36 & -0.2 & 0.27 & 971 & 452 \\
		FREI & 301.019 & -62.194 & 0.66 & -0.2 & 0.05 & 348 & 348 \\
		FTP4 & 162.565 & -78.928 & 0.15 & -0.07 & -0.09 & 283 & 447 \\
		GLDK & 259.412 & -72.233 & 1.55 & -0.9 & 0.14 & 649 & 572 \\
		GMEZ & 291.464 & -73.885 & 0.94 & 0.17 & 0.66 & 897 & 745 \\
		HAAG & 281.713 & -77.038 & 0.9 & 0.11 & 0.43 & 877 & 765 \\
		HOWE & 210.567 & -87.416 & 0.63 & 0.03 & -0.1 & 96 & 664 \\
		HOWN & 273.233 & -77.528 & 1.93 & 0.41 & 0.92 & 684 & 589 \\
		HTON & 298.269 & -74.08 & 0.13 & -0.04 & 0.09 & 61 & 1015 \\
		HUGO & 294.332 & -64.963 & 1.07 & -0.05 & -0.31 & 235 & 243 \\
		IGGY & 156.25 & -83.307 & 0.57 & -0.03 & -0.07 & 137 & 914 \\
		INMN & 261.12 & -74.821 & 9.39 & -2.22 & -0.09 & 357 & 274 \\
		JNSN & 293.898 & -73.077 & 0.08 & -0.14 & 0.54 & 14 & 854 \\
		LNTK & 286.097 & -74.835 & 1.39 & 0.22 & 0.37 & 875 & 693 \\
		LPLY & 269.701 & -73.111 & 2.91 & -0.92 & 0.09 & 658 & 601 \\
		LWN0 & 152.732 & -81.346 & 0.48 & -0.09 & -0.0 & 51 & 612 \\
		LXAA & 23.346 & -71.947 & -1.47 & 0.03 & 0.08 & 524 & 516 \\
		MAIT & 11.736 & -70.766 & -0.86 & 0.3 & -0.02 & 636 & 545 \\
		MAW1 & 62.871 & -67.605 & -0.46 & 0.16 & -0.18 & 656 & 632 \\
		MBIO & 303.377 & -64.24 & 0.94 & -0.07 & 0.3 & 269 & 271 \\
		MCAR & 215.696 & -76.322 & 1.36 & 0.01 & -0.54 & 570 & 759 \\
		MCM4 & 166.669 & -77.838 & 0.26 & -0.02 & -0.11 & 358 & 383 \\
		MCRG & 265.354 & -73.668 & 3.56 & -1.44 & 0.37 & 572 & 491 \\
		MIN0 & 167.164 & -78.65 & 0.25 & -0.04 & -0.09 & 356 & 503 \\
		MKIB & 294.397 & -75.276 & 0.21 & -0.02 & 0.25 & 8 & 892 \\
		MRTP & 244.898 & -74.18 & 5.42 & -1.82 & -0.48 & 515 & 446 \\
		MTAK & 247.2 & -76.315 & 10.29 & 1.86 & -0.78 & 335 & 228 \\
		OHI2 & 302.099 & -63.321 & 1.2 & -0.32 & 0.15 & 267 & 272 \\
		PALM & 295.949 & -64.775 & 1.78 & -0.16 & -0.59 & 157 & 164 \\
		PATN & 204.977 & -78.03 & 0.9 & -0.04 & -0.26 & 499 & 846 \\
		PECE & 291.444 & -85.612 & 0.86 & 0.14 & 0.11 & 173 & 402 \\
		PIRT & 274.857 & -81.103 & 1.0 & 0.26 & 0.22 & 848 & 837 \\
		PRPT & 294.661 & -66.007 & 0.99 & 0.13 & -0.21 & 372 & 256 \\
		RAMG & 178.047 & -84.338 & 0.25 & -0.01 & 0.01 & 781 & 753 \\
		RMBO & 293.606 & -83.873 & 0.89 & -0.0 & 0.08 & 202 & 299 \\
		ROB4 & 163.19 & -77.034 & 0.15 & -0.03 & -0.09 & 270 & 340 \\
		ROBN & 300.555 & -65.246 & 1.94 & 0.32 & 0.57 & 137 & 132 \\
		ROTB & 291.873 & -67.571 & 0.48 & -0.04 & 0.1 & 744 & 87 \\
		ROTH & 291.874 & -67.571 & 0.48 & -0.04 & 0.1 & 744 & 87 \\
		SCTB & 166.758 & -77.849 & 0.27 & -0.02 & -0.1 & 360 & 384 \\
		SDLY & 234.025 & -77.135 & 2.79 & 0.56 & -0.58 & 688 & 607 \\
		SLTR & 246.12 & -75.098 & 21.83 & -4.75 & -1.71 & 219 & 212 \\
		SMR5 & 292.897 & -68.13 & 0.09 & -0.15 & 0.24 & 21 & 576 \\
		SPGT & 298.948 & -64.295 & 3.97 & -1.19 & -0.01 & 91 & 86 \\
		STEW & 273.753 & -84.187 & 0.96 & 0.1 & 0.05 & 558 & 329 \\
		SUGG & 287.82 & -75.281 & 0.9 & 0.17 & 0.41 & 921 & 759 \\
		SVEA & 348.775 & -74.576 & 0.1 & -0.13 & -0.07 & 850 & 95 \\
		SYOG & 39.584 & -69.007 & -1.08 & 0.28 & 0.08 & 576 & 513 \\
		THUR & 262.44 & -72.53 & 1.91 & -0.95 & 0.06 & 639 & 579 \\
		TNB1 & 164.103 & -74.699 & -0.0 & -0.06 & -0.07 & 169 & 124 \\
		TOMO & 245.338 & -75.802 & 15.32 & 1.85 & -2.04 & 261 & 205 \\
		TRVE & 292.445 & -69.989 & 1.04 & 0.37 & 0.06 & 691 & 89 \\
		VESL & 357.158 & -71.674 & -2.21 & 0.17 & -0.02 & 239 & 352 \\
		VL01 & 169.725 & -72.45 & 1.17 & 0.51 & 0.08 & 48 & 125 \\
		VL12 & 163.727 & -72.274 & 0.26 & 0.1 & -0.04 & 878 & 363 \\
		VL30 & 162.525 & -70.599 & 1.82 & -0.15 & -0.06 & 105 & 27 \\
		VNAD & 295.746 & -65.246 & 1.7 & 0.06 & -0.56 & 171 & 184 \\
		WHN0 & 154.22 & -79.846 & 0.23 & -0.06 & -0.01 & 782 & 635 \\
		WHTM & 255.607 & -82.683 & 0.89 & 0.21 & -0.13 & 858 & 842 \\
		WILN & 279.442 & -80.04 & 0.91 & 0.22 & 0.29 & 786 & 778 \\
		WLCH & 296.18 & -70.729 & -0.07 & 0.11 & 0.07 & 58 & 402 \\
		WLCT & 272.607 & -85.369 & 0.98 & 0.16 & 0.05 & 861 & 631 \\
		WWAY & 331.597 & -81.577 & 0.25 & -0.09 & 0.16 & 869 & 715 \\
		ZHON & 76.37 & -69.371 & -0.23 & 0.12 & 0.01 & 174 & 187 \\

	\end{longtable}
\end{center}

\begin{center}
	\begin{longtable}{|l	|	c	|	c	|	c	|}
		\caption[Weights for LSKs centered at each GNSS site in the vertical, south, and east components used to isolate the elastic deformation due to mass loss of Pine Island Glacier  e.g., Figure~6 of the main text.]{Weights for LSKs centered at each GNSS site in the vertical, south, and east components used to isolate the elastic deformation due to mass loss of Pine Island Glacier  e.g., Figure~6 of the main text.} \label{tab:wLSK_PIG_coeffs} \\
		
		\hline \multicolumn{1}{|c|}{\textbf{Site}} &
		\multicolumn{1}{c|}{\textbf{C$_U$}} &
		\multicolumn{1}{c|}{\textbf{C$_S$}} &
		\multicolumn{1}{c|}{\textbf{C$_E$}} \\
		\hline 
		\endfirsthead
		
		\multicolumn{4}{c}%
		{{\bfseries \tablename\ \thetable{} -- continued from previous page}} \\
		\hline \multicolumn{1}{|c|}{\textbf{Site}} &
		\multicolumn{1}{c|}{\textbf{C$_U$}} &
		\multicolumn{1}{c|}{\textbf{C$_S$}} &
		\multicolumn{1}{c|}{\textbf{C$_E$}} \\
		\hline 
		\endhead
		
		\hline \multicolumn{4}{|r|}{{Continued on next page}} \\ \hline
		\endfoot
		
		\hline \hline
		\endlastfoot
		
		INMN &	 1.00	&	1.00	&	1.00	 \\
		BACK & 6.36$\times$10$^{-1}$ &	-4.26$\times$10$^{-1}$ &	-1.00 \\
		SLTR & -7.28$\times$10$^{-4}$ & -1.61$\times$10$^{-3}$ &	-2.61$\times$10$^{-4}$ \\
		BERP & 5.59$\times$10$^{-3}$ & -3.60$\times$10$^{-2}$ &	3.46$\times$10$^{-4}$ \\
		TOMO & 3.54$\times$10$^{-5}$ & 1.57$\times$10$^{-3}$ & 6.97$\times$10$^{-5}$ \\
		MTAK & -3.45$\times$10$^{-3}$ & 3.28$\times$10$^{-3}$ & 3.10$\times$10$^{-2}$ \\
		MRTP & 0.00 & 6.75$\times$10$^{-2}$ & -1.40$\times$10$^{-1}$ \\
	\end{longtable}
\end{center}

\begin{figure}
	\centering
	\includegraphics[width=1\textwidth]{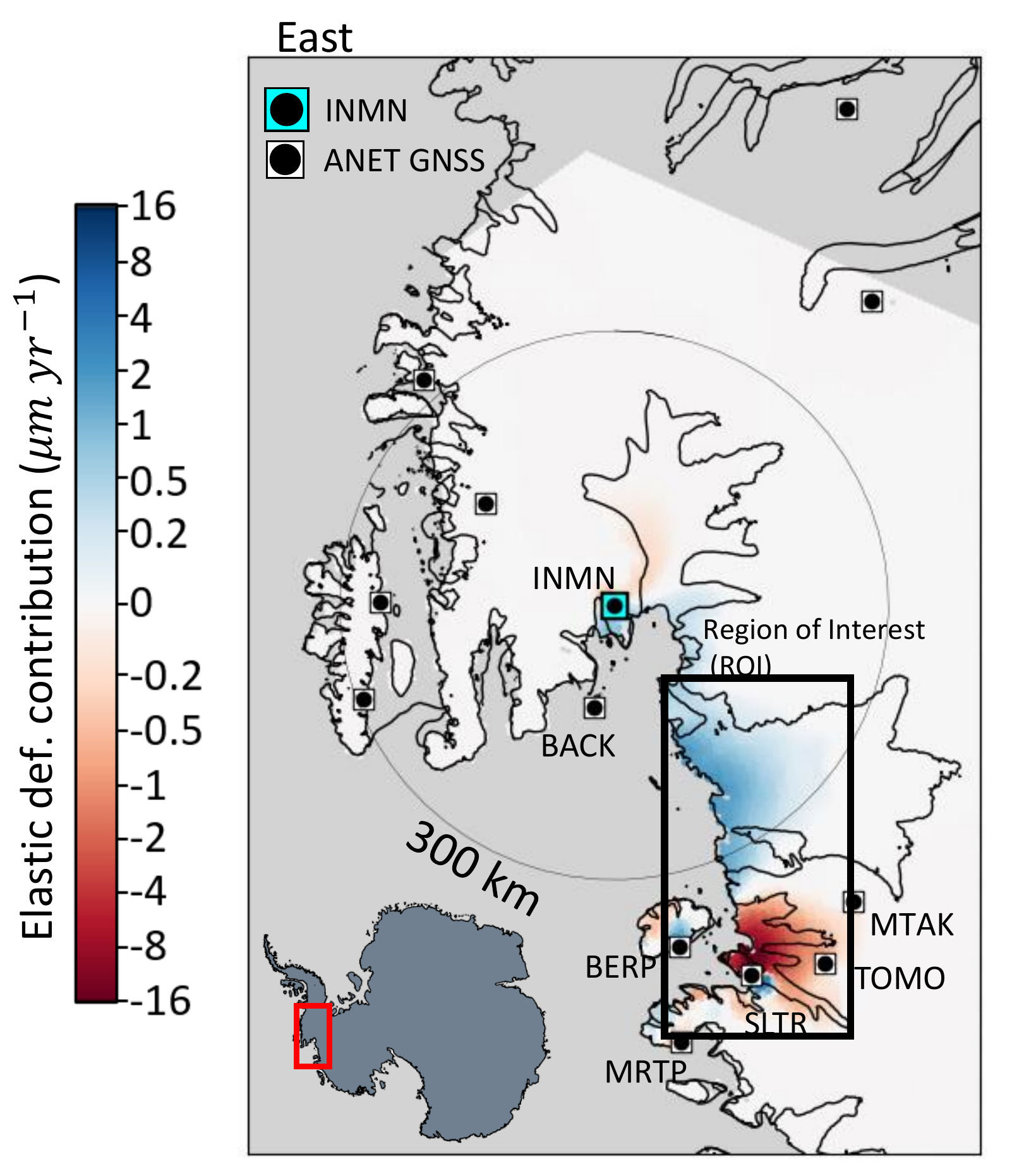}
	\caption{Weighted LSK series optimized to isolate the elastic deformation due to ice mass loss inside the outlined Region of Interest. This weighted LSK series expresses the east component of elastic deformation due to the mass loss near the grounding line of Thwaites Glacier relative to the mass loss of the PSK glaciers. The weights of each LSK are included in Table S3.}
	\label{fig:wLSKS_Thwaites}
\end{figure}

\begin{center}
	\begin{longtable}{|l	|	c	|	c	|	c	|}
		\caption[Weights for LSKs centered at each GNSS site in the vertical, south, and east components used to isolate the east component of elastic deformation due to mass loss near the grounding line of Thwaites Glacier relative to the mass loss of the PSK glaciers, e.g., Figure~S\ref{fig:wLSKS_Thwaites}.]{Weights for LSKs centered at each GNSS site in the vertical, south, and east components used to isolate the east component of elastic deformation due to mass loss near the grounding line of Thwaites Glacier relative to the mass loss of the PSK glaciers, e.g., Figure~S\ref{fig:wLSKS_Thwaites}.} \label{tab:wLSK_Thwaites} \\
		
		\hline \multicolumn{1}{|c|}{\textbf{Site}} &
		\multicolumn{1}{c|}{\textbf{C$_U$}} &
		\multicolumn{1}{c|}{\textbf{C$_S$}} &
		\multicolumn{1}{c|}{\textbf{C$_E$}} \\
		\hline 
		\endfirsthead
		
		\multicolumn{4}{c}%
		{{\bfseries \tablename\ \thetable{} -- continued from previous page}} \\
		\hline \multicolumn{1}{|c|}{\textbf{Site}} &
		\multicolumn{1}{c|}{\textbf{C$_U$}} &
		\multicolumn{1}{c|}{\textbf{C$_S$}} &
		\multicolumn{1}{c|}{\textbf{C$_E$}} \\
		\hline 
		\endhead
		
		\hline \multicolumn{4}{|r|}{{Continued on next page}} \\ \hline
		\endfoot
		
		\hline \hline
		\endlastfoot
	INMN &	--	&	--	&	1.00$\times$10$^{-1}$	\\
	BACK &	--	&	--	&	1.00$\times$10$^{-1}$	\\
	SLTR &	--	&	--	&	2.00$\times$10$^{-1}$	\\
	BERP &	--	&	--	&	-1.00	\\
	TOMO &	--	&	--	&	1.01$\times$10$^{-2}$	\\
	MTAK &	--	&	--	&	-2.53$\times$10$^{-2}$	\\
	MRTP &	--	&	--	&	4.00$\times$10$^{-1}$	\\
	\end{longtable}
\end{center}

%


%
%


\title{Supporting Information for "Insert Title"}
%
%

%
%



\authors{=Authors=}


\affiliation{=number=}{=Affiliation Address=}



%
%

%

\begin{article}

%
%

\noindent\textbf{Contents of this file}
\begin{enumerate}
\item Text S1 to Sx
\item Figures S1 to Sx
\item Tables S1 to Sx
\end{enumerate}
\noindent\textbf{Additional Supporting Information (Files uploaded separately)}
\begin{enumerate}
\item Captions for Datasets S1 to Sx
\item Captions for large Tables S1 to Sx (if larger than 1 page, upload as separate excel file)
\item Captions for Movies S1 to Sx
\item Captions for Audio S1 to Sx
\end{enumerate}

\noindent\textbf{Introduction}


\noindent\textbf{Text S1.}
%


\noindent\textbf{Data Set S1.} 


\noindent\textbf{Movie S1.} 


\noindent\textbf{Audio S1.} 
\end{article}
\clearpage


%
%
%
%
%
%
%